# Database of 2D hybrid perovskite materials: open-access collection of crystal structures, band gaps and atomic partial charges predicted by machine learning

Ekaterina I. Marchenko [1,3,#], Sergey A. Fateev [1,#], Andrey A. Petrov [1], Vadim V. Korolev [2,4], Artem A. Mitrofanov [2,4], Andrey V. Petrov[5], Eugene A. Goodilin [1,2], Alexey B. Tarasov [*1,2]

[1] Laboratory of New Materials for Solar Energetics, Faculty of Materials Science, Lomonosov Moscow State University; 1 Lenin Hills, 119991, Moscow, Russia

[2] Department of Chemistry, Lomonosov Moscow State University; 1 Lenin Hills, 119991, Moscow, Russia

[3] Department of Geology, Lomonosov Moscow State University; 1 Lenin Hills, 119991, Moscow, Russia

[4] Science Data Software, LLC, 14909 Forest Landing Cir, Rockville, MD 20850, United States

[5] Institute of Chemistry, Saint-Petersburg State University, 26 Universitetskii prospect, 198504, Saint-Petersburg, Russia

**ABSTRACT:** We describe a first open-access database of experimentally investigated hybrid organic-inorganic materials with two-dimensional (2D) perovskite-like crystal structure. The database includes 515 compounds, containing 180 different organic cations, 10 metals (Pb, Sn, Bi, Cd, Cu, Fe, Ge, Mn, Pd, Sb) and 3 halogens (I, Br, Cl) known so far and will be regularly updated. The database contains a geometrical and crystal chemical analysis of the structures, which are useful to reveal quantitative structure-property relationships for this class of compounds. We show that the penetration depth of spacer organic cation into the inorganic layer and M-X-M bond angles increase in the number of inorganic layers (n). The machine learning model is developed and trained on the database, for the prediction of a band gap with accuracy within 0.1 eV. Another machine learning model is trained for the prediction of atomic partial charges with accuracy within 0.01 e. We show that the predicted values of band gaps decrease with an increase of the n and with an increase of M-X-M angles for single-layered perovskites. In general, the proposed database and machine learning models are shown to be useful tools for the rational design of new 2D hybrid perovskite materials.

## Introduction

The family of organic-inorganic layered structures, often referred as "2D hybrid perovskites", is derived from the perovskite structural type, exhibits unprecedented structural flexibility which opens prospects for the design of various innovative materials for photovoltaics and optoelectronics. This class of materials shows a set of unique functional properties, such as record-breaking yield of photo- and electroluminescence, broad white-light emission, tunable narrow emission, excellent photoconductivity. [1,2]

2D hybrid halide perovskites can be regarded as a product of cutting the 3D parent compound along a specific crystallographic plane[3] as reflected by the general formula of $(A`)_{2/q}A_{n-1}B_nX_{3n+1}$, where $[A`]^{q+}$ represents singly (q=1) or doubly (q=2) charged organic spacer cation, $A^+$ is a small singly charged cation (such as $Cs^+$, $CH_3NH_3^+$, $[HC(NH_2)_2]^+$); $B^{2+} = Pb^{2+}, Ge^{2+}, Sn^{2+}$, etc.; $X^- = Cl^-, Br^-, I^-$; n is the number of layers of corner-shared octahedra within a perovskite slab.

Prediction of material properties based on crystal structures has come to be a useful approach for the directed rational design of the materials. Factual datasets, especially crystallographic databases, are an important tool for structural design since they deliver primary information needed for further analysis. The widespread use of high-throughput density functional theory (HT-DFT) calculations in materials science has contributed to the emergence of extensive databases[4–8] containing optimized crystal structures and the values of the most important physicochemical properties. Such projects stimulate the discovery of fundamentally new materials and the use of already known compounds in new areas. Recently, a data-driven approach has been fruitfully used to extract quantitative structure-property relationships (QSPR) hidden in publicly available materials repositories[9,10]. The screening of available datasets for specific applications was significantly accelerated by rapidly developing machine learning (ML) models[11–19]. Despite the growing interest in 2D hybrid perovskites, there is still no specific database for these structures.

In this work, we present for the first time an open-access regularly updated database of 2D hybrid perovskites containing experimental values of the band gap and structural parameters as well as a number of structural descriptors with charges and band gap values calculated by machine learning approach which can be used to derive correlations "chemical composition – structure – property". Currently, the database includes 515 compounds composed of halometallate framework consisting of cations $M^{n+}$ of different metals (including Pb, Sn, Bi, Cd, Cu, Fe, Ge, Mn, Pd, Sb) and halide anions $X^-$ ($Cl^-$, $Br^-$, $I^-$) as a central atom and ligands of $[MX_6]$ octahedron respectively and 180 different organic spacer cations (protonated nitrogen- or phosphorus-containing organic bases) occupying cuboctahedral voids or interlayer space. In sum, the database contains 414 (100) layered perovskites, 27 – (110) layered perovskites, 7 – (111) layered perovskites and 67 structures that are not ascribed to a specific structure derived from the perovskite structural type and containing metal-halide octahedral frameworks with different topologies. There are (100) structures with different number of octahedra layers (n), including 310 items with n = 1 (single-layered), and multilayered structures: 50 items with n = 2, 34 items with n = 3, 11 items with n = 4, 3 items with n = 5, 1 item with n = 6, and 2 items with n = 7. Among (100) structures, iodide-compounds dominate (236 records) over bromide (103) and chloride (73) compounds. A progressive extension of the dataset is expected as compounds with new organic cations continue to emerge.

**Methods**

The present database of 2D hybrid perovskites contains both the information about the crystal structure of the materials (like the first-level databases[20]) and different parameters of these materials calculated from our crystal chemical analysis and predicted from machine learning models.

The structural information such as chemical formulas of compounds, space groups of symmetry, the number of octahedra layers (n), types of organic cation was extracted from *.cif files of experimentally investigated compounds supplementing the cited articles or from crystallographic data presented in another form. In the latter case, the *.cif files were manually created for convenience and for further data analysis and processing. Visualization of crystal structures was carried out using the VESTA program[21].

Topological and crystal chemical analysis

Having performed a topological and crystal-chemical analysis of the collected structures (*.cif), we derived the following parameters: the M–X bond length, X–M–X and M–X–M bond angles and penetration depth of spacer organic cations into cuboctahedral voids of the layer of corner-shared octahedral ("perovskite layers").

For individual representative series of related compounds (such as the series of (100) layered perovskites with the same spacer cation and different n), such parameters as volumes distortion of $BX_6$ octahedra and Voronoi-Dirichlet polyhedra of organic cations were calculated.

These parameters and volumes of Voronoi-Dirichlet polyhedra of organic cations which characterize relative atomic size in crystal structures are obtained from a geometrical analysis of the crystal structures from dataset performed using the ToposPro program package[22] (see SI).

To determine the degree of distortion of $BX_6$ octahedra, we used the equation introduced by Alonso et al.,[23] and commonly used for evaluation of distortion degree of layered perovskites

$$\Delta d = \frac{1}{6} \sum \left[\frac{d_n - d}{d}\right]^2, \qquad (1)$$

where $d_n$ - individual B-X distances, d - arithmetic mean values of the individual B-X distances.

The values of penetration depth of spacer organic cations into the inorganic layer of corner-shared octahedra (hereinafter – "penetration") were calculated for each structure with integer occupancies of N and X atoms according to the methodology described in [24]. The penetration depth was defined as the average distance between the protonated nitrogen atoms of the spacer organic cation and the plane of axial halogen atoms of the perovskite slab.

Development of machine learning models

First machine learning model (MLM1) for predicting the band gap values was developed and trained on the reliable set of 2D hybrid perovskite structures from the database and corresponding band gap values calculated by DFT. The electronic structure was calculated by DMol3 module of Materials Studio software package [25,26] using the atomic basis DNP+ (Double Numerical plus polarization with the addition of diffuse functions) with spin-orbital coupling. The calculated values of the band gaps are in a good agreement with the experimental data (Figure 1) and therefore was used as a training set for MLM1.

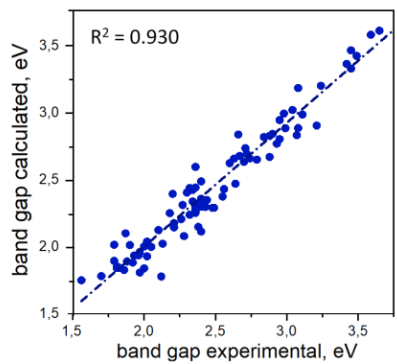

Figure 1. Band gaps (theoretical (DFT) vs experimental) of structures from the train set. Only the structures with reported experimental values of the band gap are shown here. The band gap values are listed in SI file.

The second machine learning model for predicting partial atomic charges (MLM2) was trained on the computation-ready, experimental metal-organic frameworks (CoRE MOFs) database [27–29]. Atomic point charges for 2932 MOFs[30] obtained with the DDEC charge partitioning method were used for the presented training/validation model.

Data representation for partial charge prediction does not require averaging since this quantity is atom-wise, i.e., it corresponds to a distinct atom. We used elemental properties and structural descriptors (Voronoi tessellation-based features) to characterize an atomic environment[31]. The suitable structure representation is a critical point for high-performance QSPR modeling[32,33]. To represent hybrid organic-inorganic perovskites for band gap prediction, we used the smooth overlap of atomic positions (SOAP) kernel[34] (more detailed information on SOAP is presented in SI). The local environment of each atom is determined as a fixed length vector. We introduced the fast average kernel over all atoms as descriptors for the crystal structures.

A detailed description of the ML pipeline for partial charge prediction is provided in ref. [35]. The gradient boosting decision tree method implemented with XGBoost library[36] was used for both endpoints. We tested the performance on an external test set (20 % and 10% of structures from the initial set) with five- and ten-fold cross-validation for band gap and partial charge prediction, respectively.

## RESULTS AND DISCUSSION

The information presented in the database for each compound is stored in open-access on www.pdb.nmse-lab.ru.

Each entry of the database contains the following information about the given compound:

1) CIF file;
2) General formula;
3) Number of octahedra layers;
4) Space group symmetry;
5) Type of layered perovskite structure (such as Ruddlesden-Popper (RP), Dion–Jacobson (DJ));
6) M-X bond lengths;
7) M-X-M and X-Pb-X angles;
8) Penetration depth of spacer organic cations into inorganic layers;
9) Reference to the original work;
10) DOI of the reference;
11) Calculated band gaps;
12) Experimental optical band gaps;
13) Partial atomic charges;

We analysed the structural parameters of compounds, which primarily determine the electronic structure of these compounds: distortions of the inorganic framework and penetration depth of spacer organic cations into inorganic layers. The distortions of the inorganic framework can be divided into two parts. The first is the distortion of the $MX_6$ octahedra ($\Delta d$) themselves (Figure 2), represented by X-Pb-X angles. The second parameter is octahedral tilting which can be determined from the M-X-M angle(s) [24]. We analyzed $\Delta d$ of different octahedral layers ($\Delta d_1$, $\Delta d_2$, $\Delta d_3$, $\Delta d_4$) as shown for some series of (100) 2D hybrid perovskites with different cations (Figure S3 SI), such as n-butylammonium ($BA^+$), octane-1,8-diammonium ($ODA^{2+}$), guanidinium ($GUA^+$) and 4-(aminomethyl)-piperidinium ($4AMP^+$) (see table S2 in SI). It was found that the octahedra of the layers neighboring with the spacer organic cations, are the most distorted ones in the multilayer structures (n > 1). To investigate further this dependency we estimated the volumes of the Voronoi–Dirichlet polyhedra ($V_{VDP}$)[37] for spacer organic cations and revealed that it decreases with an increase of the $\Delta d_1$ (Fig. S4 in SI, Table S3).

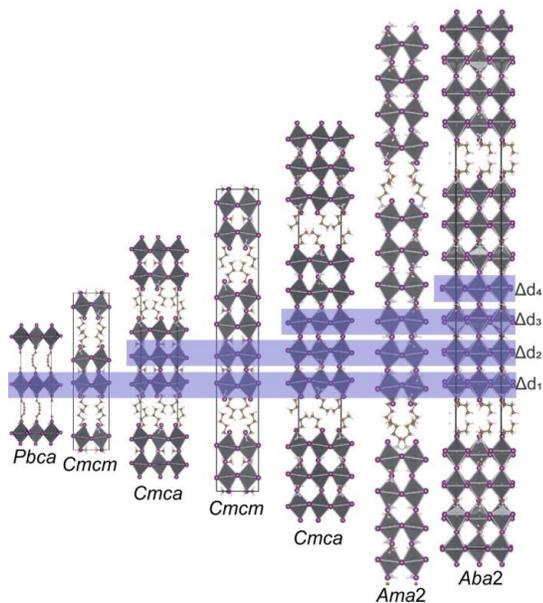

Figure 2. Series of 2D lead iodide perovskite structures (n=1-7) with butylammonium (BA$^+$) as a spacer organic cation BA$_2$(CH$_3$NH$_3$)$_{1-n}$Pb$_n$I$_{3n+1}$. The layers of octahedra with corresponding distortions Δd$_{1-4}$ are highlighted. Carbon, hydrogen, nitrogen, and iodine atoms are shown in dark brown, light pink, gray, and purple, respectively. PbI$_6$ octahedra are shown in gray.

The increase of another distortion parameter, the average M-X-M bond angle, towards an ideal 180° indicates a lower perovskite layer tilting. This should lead to the decrease of the band gap with the decreasing of the tilting for the series of homological 100 perovskites with different spacer organic cations. Indeed, we show that the predicted band gap values decrease with increasing of M-X-M angles (Figure 3).

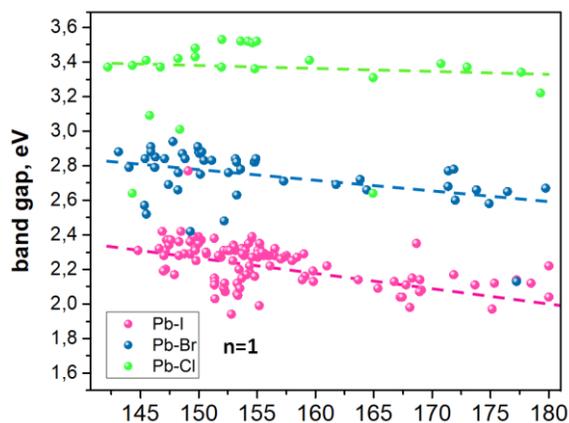

Figure 3. The dependence of the calculated band gap by ML on the Pb-X-Pb angles for single-layered (100) compounds (A`)$_{2/q}$A$_{n-1}$B$_n$X$_{3n+1}$ with different spacer organic cations [A`]$^{q+}$ from the database. The dotted lines are given for eye guidance only.

In order to couple the parameters of an inorganic framework with the features of organic cations we calculate the values of penetration depths of spacer organic cations into inorganic layers. It was shown for 10 series of (100) perovskites with n = 1-4 and Pb-I inorganic framework with different organic cations that with an increase in the number of layers, the penetration and the deviation angle of Pb-X-Pb bonds increase (Figures 4a and 4b). An increase in the penetration with an increase of n can be explained from an electrostatic point of view. Indeed, with an increase of n, the dielectric constant of the perovskite slab increase[2]. Consequently, the attraction between such a slab and an "electric double layer" formed by the charged planes of organic cations (+) and terminal halides (-) should also enhance, resulting in greater mutual penetration of the organic cation and the layer. The large scatter in the values of penetration for structures with the same n is simply explained by the simultaneous presence in the database of different structures for one compound including those refined at different temperatures.

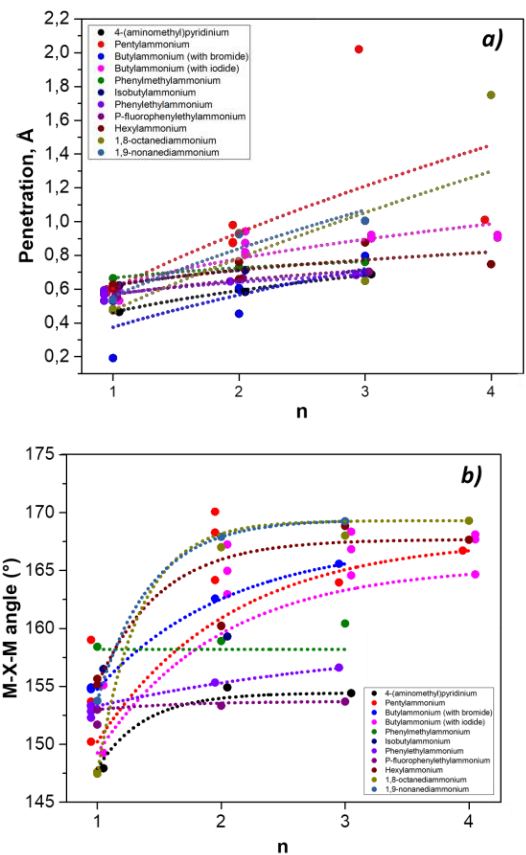

Figure 4. The penetration depths of spacer organic cation (a) and M-X-M bond angles (b) for series of 2D lead halide perovskites (A`)$_{2/q}$A$_{n-1}$B$_n$X$_{3n+1}$ with different spacer organic cations [A`]$^{q+}$ listed in the legend for n = 1-4. The points are offset along the x axis from the n values for clarity.

Moreover, we found a correlation between the composition of the organic cation and its penetration depth into the structure for a number of compounds with the same inorganic framework. So, for (100) single-layered compounds structures with Pb-I(Br) and Sn-I inorganic frameworks the lowest values of penetration depth is observed for spacer cations with relatively short carbon chain containing heteroatoms (see Figures S5, S6, S9 and S11).

To show the potential of the dataset information for prediction the parameters relevant for the rational design of hybrid perovskite materials we implemented two machine learning models (MLM): MLM1 for band gap (essential for photovoltaic applications) calculation and MLM2[35] for atomic partial charges (necessary for the development of force fields models for high-throughput simulations).

Using our MLM1 algorithm, we predicted the band gap for all compounds of our database without partial occupations of M and X crystallographic positions with good accuracy: the root mean square deviation (RMSD) and mean absolute deviation (MAD) were 0.136 and 0.103 eV, respectively (Figure 5). Giving a smaller mismatch, our machine learning model outperforms another similar model predicting band gap values for organic crystal structures available to date[38].

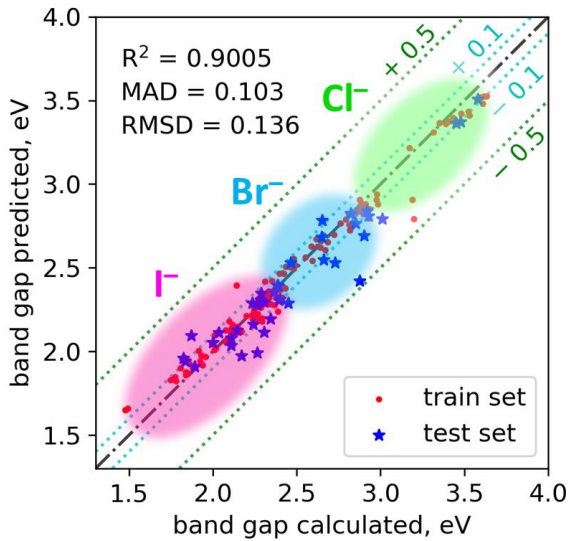

Figure 5. Band gaps of 2D hybrid halide perovskites calculated by DFT and predicted by MLM1 model trained on the reliable set of the structures from the database.

Importantly, our machine learning model predicts a clear trend of decreasing of the band gaps with an increase of n for all known series of 2D lead halide perovskites compounds $(A`)_{2/q}A_{n-1}B_nX_{3n+1}$ with different spacer organic cations (Figure 6). The predicted trend fairly matches with a fundamental dependency of band gaps with increasing of the number of layers[2]. This implies that the band gaps for multilayer perovskites having the highest promise for a photovoltaic application can be predicted correctly. In addition, it can be seen that the slope of trends of band gap from n significantly depends on the tilting of octahedra and type of spacer cation. We assume that this dependency originates from the differences in M-X-M bond angles for various series of compounds in Figure 6. Such correlations have key importance for fine-tuning of the optoelectronic properties of the layered perovskite materials.

We believe that the proposed high-throughput machine learning approach has many potential applications and can be especially useful, for instance, for band gap adjustment of light absorbing materials used in tandem solar cells.

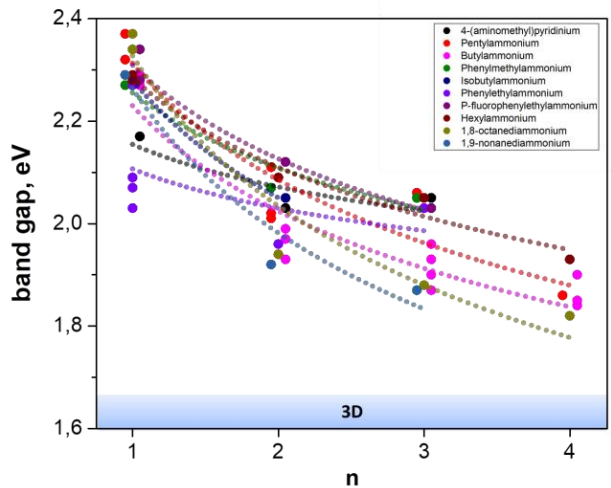

Figure 6. Band gaps calculated by MLM1 model for series of 2D hybrid halide perovskites $(A`)_{2/q}A_{n-1}B_nX_{3n+1}$ with different spacer organic cations $[A`]^{q+}$ listed in the legend for n = 1-4. The points are offset along the x axis from the n values for clarity.

## CONCLUSIONS

We have constructed the database of 515 layered hybrid perovskites compounds. Based on the crystal chemical analysis we found several novel structural correlations. Particularly, we found that the $V_{VDP}$ of spacer organic cation in the structure decreases with increasing distortion of neighboring octahedra. We show for 10 series of (100) perovskites with n = 1-4 with different spacer organic cations that with an increase in the number of layers, the penetration and the deviation angle of Pb-X-Pb bonds increase.

We developed a machine learning model for band gap calculating and successfully trained it on the dataset of our database. The proposed model predicts correctly the fundamental trend of decreasing band gap values with an increase of the number of inorganic layers (n), which is essential for tuning the band gap from 3D structures to multi-layered or monolayered structures. A successful application of the de-veloped ML model on the experimental structural data of 2D hybrid halide perovskites collected in our database opens prospect for high-throughput searching of essential quantitative structure-property relationships (QSPR) for layered hybrid halide perovskites. We believe that the newly revealed QSPRs will become a useful tool for development of effective photovoltaic and optoelectronic devices based on hybrid perovskites.

## ASSOCIATED CONTENT

The Supporting Information is available free of charge via the Internet at http://pubs.acs.org.

The database is available free of charge on http://pdb.nmse-lab.ru

The interface for band gap predicting is available on https://eg.scidatasoft.com/

The algorithm for calculation of effective charges is available on https://mof.scidatasoft.com/

## AUTHOR INFORMATION

### Corresponding Author

* alexey.bor.tarasov@yandex.ru

### Author Contributions

#These authors contributed equally. The manuscript was written through contributions of all authors. All authors have given approval to the final version of the manuscript.

### Notes

The authors declare no competing financial interest.

## ACKNOWLEDGMENT

This work was financial supported by a grant from the Russian Science Foundation, project number 19-73-30022.
The research is carried out using the equipment of the shared research facilities of HPC computing resources at Lomonosov Moscow State University.
The DFT calculations were carried out using computational re-sources provided by Resource Center "Computer Center of SPbU"

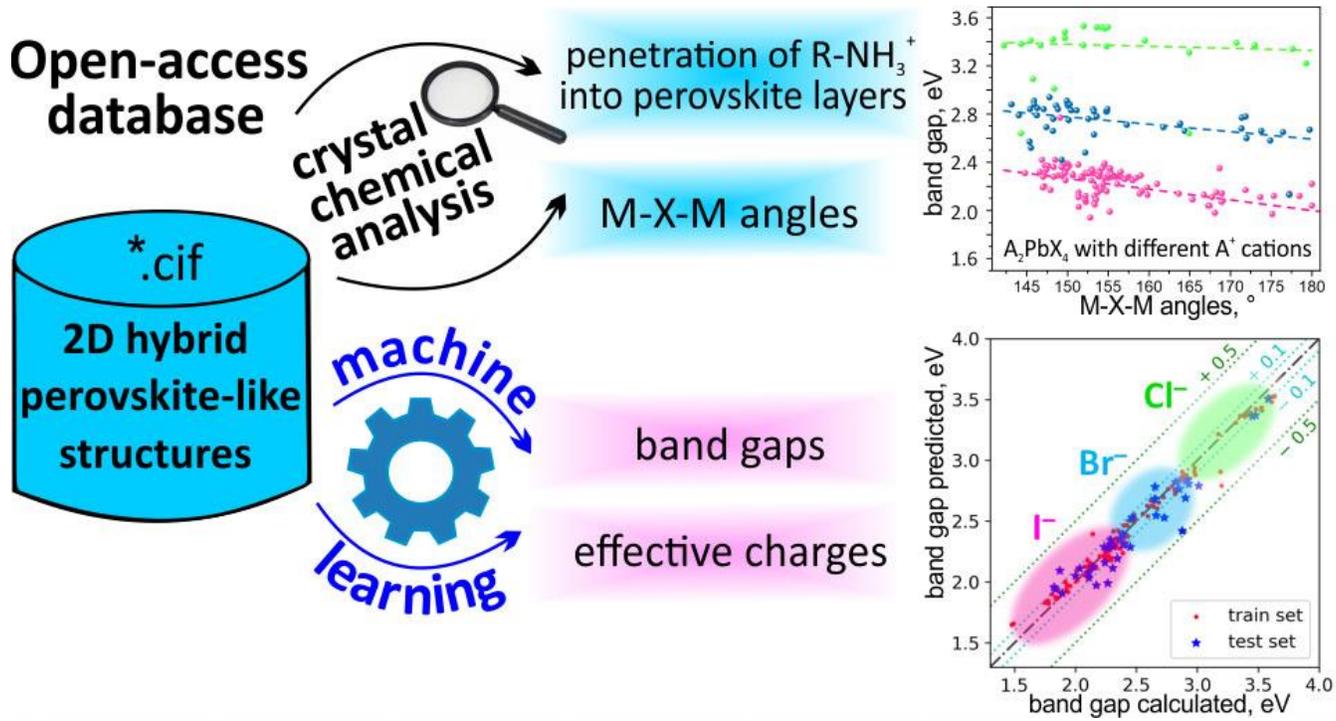
8